\documentclass[tighten, twocolumn, trackchanges]{aastex61}
\usepackage{bm, graphicx, subfigure, amsmath, morefloats, mathtools}
\hypersetup{allcolors = [rgb]{0., 0., 0.5}} 
\usepackage[toc,page]{appendix}

\usepackage{longtable}

\newcommand{\xone}{\ensuremath{x_n}}
\newcommand{\xtwo}{\ensuremath{x_{n'}}}

\providecommand{\mas}{\ensuremath{\rm mas}}
\providecommand{\masyr}{\ensuremath{\rm mas.yr}^{-1}}

\providecommand{\pc}{\ensuremath{\rm pc}}

\providecommand{\Gyr}{\ensuremath{\rm Gyr}}
\providecommand{\Myr}{\ensuremath{\rm Myr}}

\providecommand{\dex}{\ensuremath{\rm dex}}
\providecommand{\au}{\ensuremath{\rm AU}}
\providecommand{\teff}{\ensuremath{{T}_{\rm eff}}}
\providecommand{\logg}{\ensuremath{\log{g}}}
\providecommand{\msun}{\ensuremath{{\rm M}_\odot}}
\providecommand{\feh}{\ensuremath{\rm [Fe/H]}}

\received{\today}
\revised{\today}
\accepted{}
\submitjournal{ApJ}

\begin{document}

\title{Precise Ages of Field Stars from White Dwarf Companions}

\author[0000-0001-9256-5516]{M. Fouesneau}
\affiliation{Max-Planck-Institut f\"ur Astronomie, K\"onigstuhl 17, D-69117 Heidelberg, Germany}

\author[0000-0003-4996-9069]{H-W.~Rix}
\affiliation{Max-Planck-Institut f\"ur Astronomie, K\"onigstuhl 17, D-69117 Heidelberg, Germany}

\author[0000-0002-5775-2866]{T.~von Hippel}
\affiliation{Max-Planck-Institut f\"ur Astronomie, K\"onigstuhl 17, D-69117 Heidelberg, Germany}
\affiliation{Physical Sciences Department, Embry-Riddle Aeronautical University, Daytona Beach, FL 32114, USA}

\author[0000-0003-2866-9403]{D.~W.~Hogg}
\affiliation{Max-Planck-Institut f\"ur Astronomie, K\"onigstuhl 17, D-69117 Heidelberg, Germany}
\affiliation{Center for Cosmology and Particle Physics, Department of Physics, New York University, 726 Broadway, New York, NY 10003}
\affiliation{Center for Data Science, New York University, 60 5th Avenue, New York, NY 10011, USA}
\affiliation{Flatiron Institute, Simons Foundation, 162 Fifth Ave, New York, NY 10010}

\author[0000-0001-9289-0589]{H~Tian}
\affiliation{Max-Planck-Institut f\"ur Astronomie, K\"onigstuhl 17, D-69117 Heidelberg, Germany}
\affiliation{China Three Gorges University, Yichang  Hubei 443002 China}
       
\correspondingauthor{M. Fouesneau}
\email{fouesneau@mpia.de}

\begin{abstract} 
Observational tests of stellar and Galactic chemical evolution call for the
joint knowledge of a star's physical parameters, detailed element abundances,
and precise age. For cool main-sequence (MS) stars the abundances of many
elements can be measured from spectroscopy, but ages are very hard to determine.
The situation is different if the MS star has a white dwarf (WD) companion and a
known distance, as the age of such a binary system can then be determined
precisely from the photometric properties of the cooling WD. As a pilot study
for obtaining precise age determinations of field MS stars, we identify nearly
one hundred candidate for such wide binary systems: a faint WD whose
GPS1 proper motion matches that of a brighter MS star in Gaia/TGAS with a good parallax ($\sigma_\varpi/\varpi\le 0.05$). We model the WD's multi-band photometry with 
the BASE-9 code using this precise distance (assumed to be common for the pair)
and infer ages for each binary system. The resulting age estimates are precise
to $\le 10\%$ ($\le 20\%$) for $42$ ($67$) MS-WD systems. Our analysis more than
doubles the number of MS-WD systems with precise distances known to date, and it
boosts the number of such systems with precise age determination by an order of
magnitude. With the advent of the Gaia DR2 data, this approach will be
applicable to a far larger sample, providing ages for many MS stars (that can
yield detailed abundances for over 20 elements), especially in the age range 2
to 8\,\Gyr, where there are only few known star clusters.
\end{abstract}


\section{Introduction}\label{sec:introduction}
The two members of a binary star systems are stars born at nearly the same time
from the material of the same element composition, but usually with different
masses. Binary stars are not only interesting in themselves but offer a wide
range of avenues to measure stellar properties and learn about stellar physics.
These opportunities include the dynamical and geometrical calibration of their
masses and radii \citep{Torres2010}, or the cross-check of age or abundance
estimates. 

Binaries are also systems where some physical characteristics (e.g. age) are far
more easily or precisely estimated from one component, while other
characteristics (e.g. element composition) are far more easily estimated from
the other one; yet they should be near-identical among them: this is in
particular the case for wide well-resolved binary systems that consist of a
main-sequence (MS) stars and a white dwarf (WD). If we have the distance, the
magnitude, the color, and the atmospheric type information for a WD, we can
precisely and accurately age-date that object \citep{Bergeron2001}, yielding
$\tau_{age}$.  This age-dating draws on well-understood WD cooling curves and
initial-final mass relations (IFMR), which have been calibrated using star
clusters \citep[e.g.][]{Salaris2009}. We can then safely assume that the MS
primary component must be co-eval, which provides us $\tau_{age}$ of this MS
field star, a quantity that would be difficult or impossible to determine
(unless the star were near the MS turn-off). For MS stars, their (photospheric)
element abundances $[\vec{X}/H]$ can be estimated straightforwardly from
spectra, at least if they are FGK stars. The binary system as a whole then
provides us with a joint estimate of temperature $\teff$, luminosity $L$,
abundances $[\vec{X}/H]$, and a precise age $\tau_{age}$, which is fundamental
input for Galactic chemical evolution studies and tests of stellar evolution.
At the moment, we have excellent parallaxes for many MS stars from Gaia DR1 TGAS
\citep{Gaia2016}, but we have good direct parallax distances for only a few WDs. 

In this work, we set out to identify previously unknown wide binaries consisting of MS primaries with good TGAS parallaxes, and common proper motion WD secondaries; those secondaries are equidistant, which gives us their luminosity, thereby enabling the age determination for the whole binary system. This is the same approach that \citet{Tremblay2017} pursued, who focused on the masses and radii of their WD sample and did not determine ages.

Exploiting WD-MS binaries is by no means the only approach to determining the ages of MS field stars 
\citep[e.g.][]{soderblom2010}. For example, for stars near the MS turn-off the precise determination of $\logg$, $\teff$, and $\feh$ constrains the age well. Further, asteroseismology \citep{Chaplin2014} 
and gyrochronology \citep{Angus2015} have been recently proven powerful tools in practice. 
But those approaches are largely restricted to stars of $\gtrsim 1\,\msun$ and yield typical age uncertainties of 30\% \citep{Chaplin2014}. For Galactic (chemical) evolution, however, consistent tracers that exist across all relevant ages (1-13~\Gyr) are crucial: on the MS that applies to stars with $\lesssim 0.8\,\msun$, where asteroseismic and gyrochronological approaches are difficult and far less tested. 
In this regime, WD-MS wide binaries may be the best way forward to reach $\sim 10\% $ age precision.

This paper is organized as follows: in Section~\ref{sec:CandidateSelection} we describe the identification of likely WD-MS binary systems that have TGAS information on the MS component; in Section~\ref{sec:Ages} we then exploit the resulting precise luminosity information of the WD to derive its cooling and overall age. 
In Section~\ref{sec:Outlook} we then discuss follow-up of our analysis and the prospects of this approach with Gaia DR2 data. 

\section{Identification of Candidate WD - MS Wide Binaries}\label{sec:CandidateSelection}

We aim to identify WD-MS wide binary candidates without using the actual luminosity (or apparent magnitude) or detailed color of the possible WD component,  as these quantities should subsequently serve as constraints on the WD's age.  We cannot also rely on only spectroscopically confirmed WDs, as this would severely limit the sample in sky-coverage and apparent (WD) magnitude. Requiring a precise parallax-based distance for at least one of the components (almost inevitably the MS star) limits us to MS stars with ``good'' parallaxes from TGAS  (we adopt relative precisions $\le 5\%$).  Possible WD companions to these stars have to be nearby on the sky ($\le 50$~arcsec), and we arbitrarily restrict these further to angular separations that correspond to $\le 10,000\,\au$ at the distance of the MS primary, $D_{MS}$. 
Any wide but gravitationally bound WD companions will be co-moving (typically within $\le 1$~km/s) in their proper motions, $\vec{\mu}$ (at separations $\Delta\theta \ll 1$~radian).  
This means that as a first step we need to identify the binary components as co-moving pairs of stars (one of them in TGAS) that are projected to within $\le 10,000\,\au$ on the sky (at $D_{MS}$).

\begin{figure}
\includegraphics[width=\columnwidth]{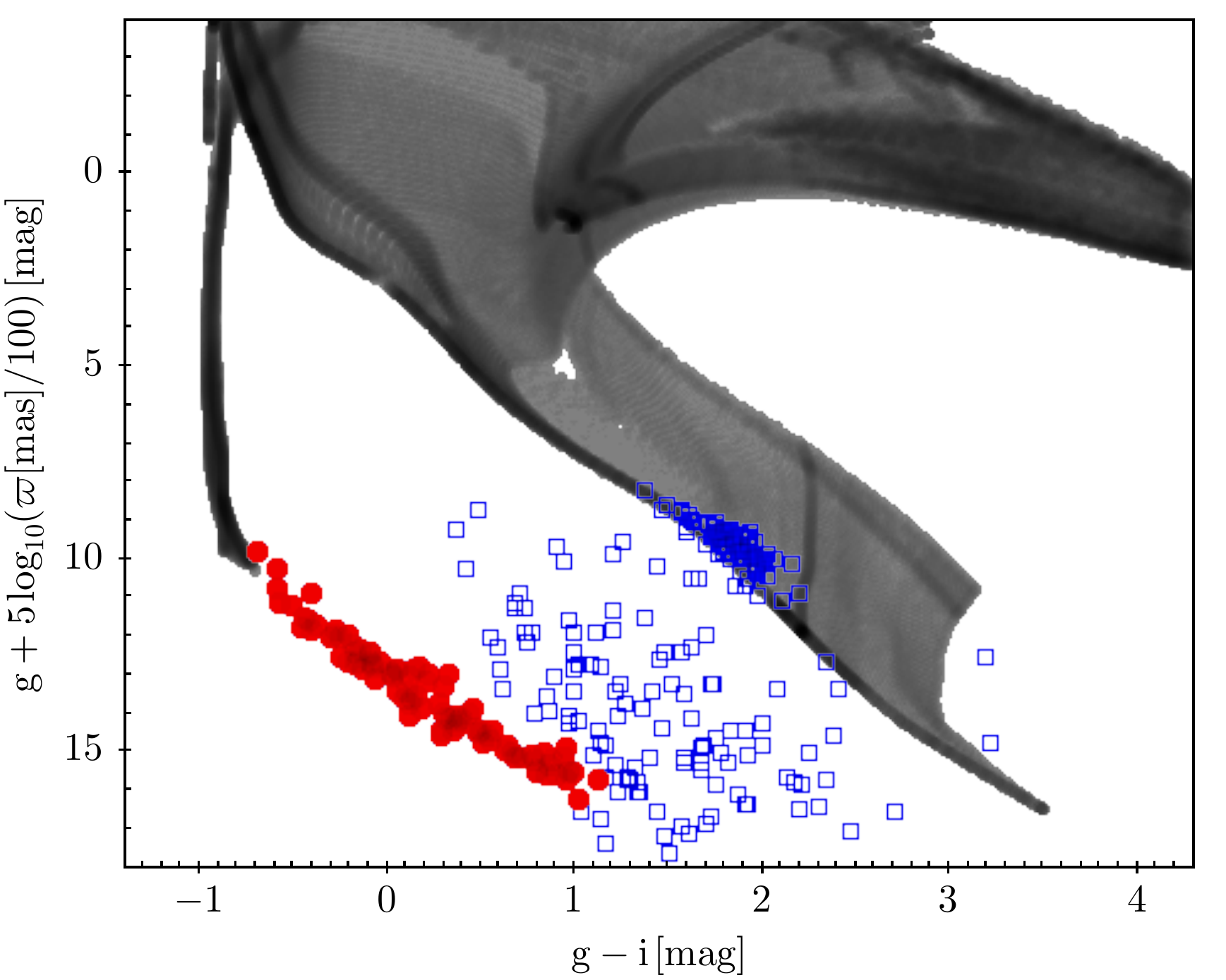} 
  \caption{Color-(absolute) magnitude diagram of candidate binary companions to TGAS primary stars (blue). The WDs analyzed in this study are highlighted in red. These candidates were selected to be within $200\,\pc$, to have separations $<10,000\,\au$, to have identical joint proper motions within 5$\sigma$. We also eliminated TGAS primaries with very small proper motions to reduce background contamination (see Appendix~\ref{appendix:gps1query}).
  Assuming the candidate secondaries to be equidistant to the TGAS primaries, we can place them on a color-magnitude diagram.
  The comparison with MESA isochrones (gray dots, \citealt{Dotter2016}) shows a clear main sequence, and a very clear WD sequence, with some remaining contaminants (that are far from any isochrone or cooling curve).  For the present paper, we only consider the candidate WD companions, identified from this diagram (red).
  (See Appendix~\ref{appendix:gps1query} for database query.)
  }
\label{fig:fig_selection}
\end{figure}

The WD secondaries will generally be much fainter than the MS primaries from
TGAS. Therefore, we cannot draw on TGAS for their proper motions. Combining
extensive sky coverage ($3\pi$) with proper motion precision and accuracy, the
GPS1 catalog \citep{Tian2017} may be the best current source of such proper
motions. 
Specifically, we queried (see Appendix~\ref{appendix:gps1query}) 
the GPS1 catalog to return the possible companions to all $\sim 100,000$ TGAS stars that had parallax measurements better than $5\%$ and parallax estimates greater than $5\,\mas$ (i.e. $<200\,\pc$ in the limit of exact parallaxes); we also required that the projected separation corresponded to less than $\le 10,000\,\au$ and that the proper motions among the potential pair were consistent at the $5\sigma$ level. We further required that the PS1 photometry for the companion was $\sigma < 0.05$~mag in $girz$, that the sources had colors consistent with the $(g-r)~vs.~(r-i)$ color-color locus of WDs. Finally, we eliminated candidates that had very wide separation, yet low proper motions, as they are particularly susceptible to (background) contamination. The specifics are detailed in Appendix~\ref{appendix:gps1query}.

This above selection left us with a wide binary sample of about 150 objects, where we expect
the companions to the TGAS MS stars to be either fainter MS stars or WDs.
Adopting the parallax-distance to the primary MS, we can construct a color --
absolute magnitude diagram for the candidate companions, which is shown in
Fig.\ref{fig:fig_selection}. It shows both a clear MS and a WD sequence,
attesting to the fact that for the most part, we have selected equidistant (and
presumably bound) companions; there are few interlopers, apparent in 
Fig. \ref{fig:fig_selection} as objects whose color-magnitude position is
inconsistent with stellar isochrones of WD cooling curves. Some of these objects
are MS-MS binaries, others may just be background contaminants. For the present
paper, we are not interested in the MS secondary components and the obvious
interlopers, so we eliminate them from further consideration.

\section{Age Constraints on the Wide Binary Systems}\label{sec:Ages}

\begin{figure}
\includegraphics[width=\columnwidth]{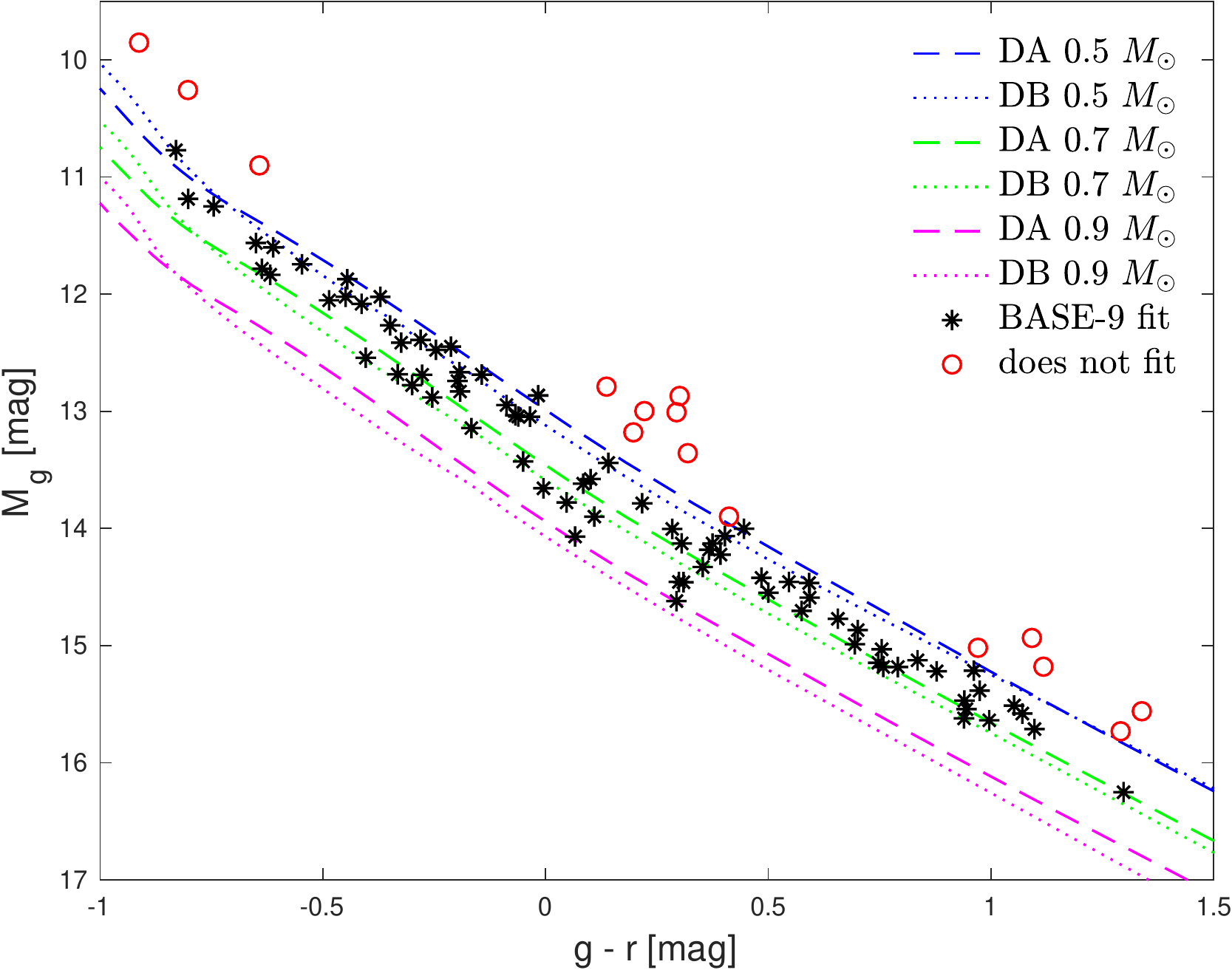} 
  \caption{Comparison of the color-(absolute) magnitude distribution of our candidate WDs to a set of cooling curves for DA (dashed; and DB, dotted) WDs with masses between $0.5$ and $0.9\,\msun$. The plot also indicates (red circles) the candidates for which BASE-9 modeling could not find acceptable solutions, presumably because they are not single WDs at the primary star's distance.}
\label{fig:fig_data_cooling_curve}
\end{figure}

We are now left with a set of 91 candidate WDs (cWD), whose distances are precisely constrained by the parallaxes to their
companions. Of those, 15 are brighter (Figure \ref{fig:fig_data_cooling_curve}, 
red circles) than the predictions from the $0.5\,\msun$
cooling curve of \citet{Bergeron1995}, which implies they have masses that are too low 
to be consistent with single-star evolution during the age of the Universe. Thus,
these objects are either the result of common envelope evolution, or are
themselves unresolved binary WDs, or the photometry is contaminated, e.g. by a
background source.  We conservatively eliminate these objects from further consideration in this preliminary work.

To now infer precisely the ages of these WDs, we need to know and compare their
trigonometric parallaxes, their spectral energy distributions (SEDs), and 
their atmospheric types (DA, DB, etc.) to models. Such modeling requires an understanding of WD cooling processes, of 
the initial-final mass relation (IFMR) of WDs, and an understanding of the precursor 
stars' lifetimes as a function of mass and metallicity.  In practice, this inference can be 
accomplished via the software suite BASE-9 \citep{vonHippel2006,DeGennaro2008,vanDyk2009,Stein2013,Stenning2016},
which fits the SED of each cWD, using the Gaia trigonometric parallax for the MS star as prior information. 

For the present context, BASE-9 serves as a flexible software package that combines 
stellar evolution models \citep[e.g.][]{dotter2008},
an IFMR \citep[e.g.][]{Salaris2009,williams2009}, WD interior cooling models 
(e.g., \citealt{Althaus1998}; \citealt{Montgomery1999}, updated and expanded for our 
use in 2011; \citealt{Renedo2010}), and WD atmosphere models \citep[e.g.][updated regularly on-line]{Bergeron1995}, with photometric constraints in a wide 
range of possible passbands. BASE-9 accounts for the individual uncertainties for all 
data; the ancillary information (e.g. parallax) and astrophysical knowledge are incorporated through 
the prior distributions.  \citet{OMalley2013} demonstrated BASE-9 
derives reliable posterior age distributions for individual field WDs and {von 
Hippel et al (2018, in prep)} show how the derived WD age precision depends on
WD masses, number and quality of photometric bands, and parallax precision.

The WD ages we derive below will indicate that these systems are most likely to
be disk or thick disk stars.  Because we do not yet have spectroscopic
abundances (of the MS primary), we set the prior distribution on metallicity to
be a broad Gaussian with a mean $\langle\feh\rangle = -0.5\,\dex$ and a
dispersion $\sigma(\feh) = 1.0\,\dex$.  While we also do not have the
line-of-sight absorption for these stars, they are all closer than $200\,\pc$,
with most being nearer than $100\,\pc$, so we set a strong prior on the
absorption of $A_{\rm 0} \approx  0$\,mag.

Using these input data and constraints, we ran BASE-9 on each cWD
individually, without further knowledge of the properties of its MS
companion, employing \citet{dotter2008} precursor models, the \citet{williams2009}
IFMR, \citet{Montgomery1999} WD interiors, and \citet{Bergeron1995}
WD atmospheres.  Without spectroscopy, we do not know which objects are
H-atmosphere (DA) WDs and which are DBs.  Fortunately for our analysis,
nature makes predominantly DA WDs ($\sim$75\%; \citealt{Tremblay2008}), and it is therefore a good
initial assumption that those cWDs that have posterior distance
probabilities consistent with their candidate MS companion Gaia parallaxes,
are indeed DAs. 

\begin{figure}
\includegraphics[width=\columnwidth]{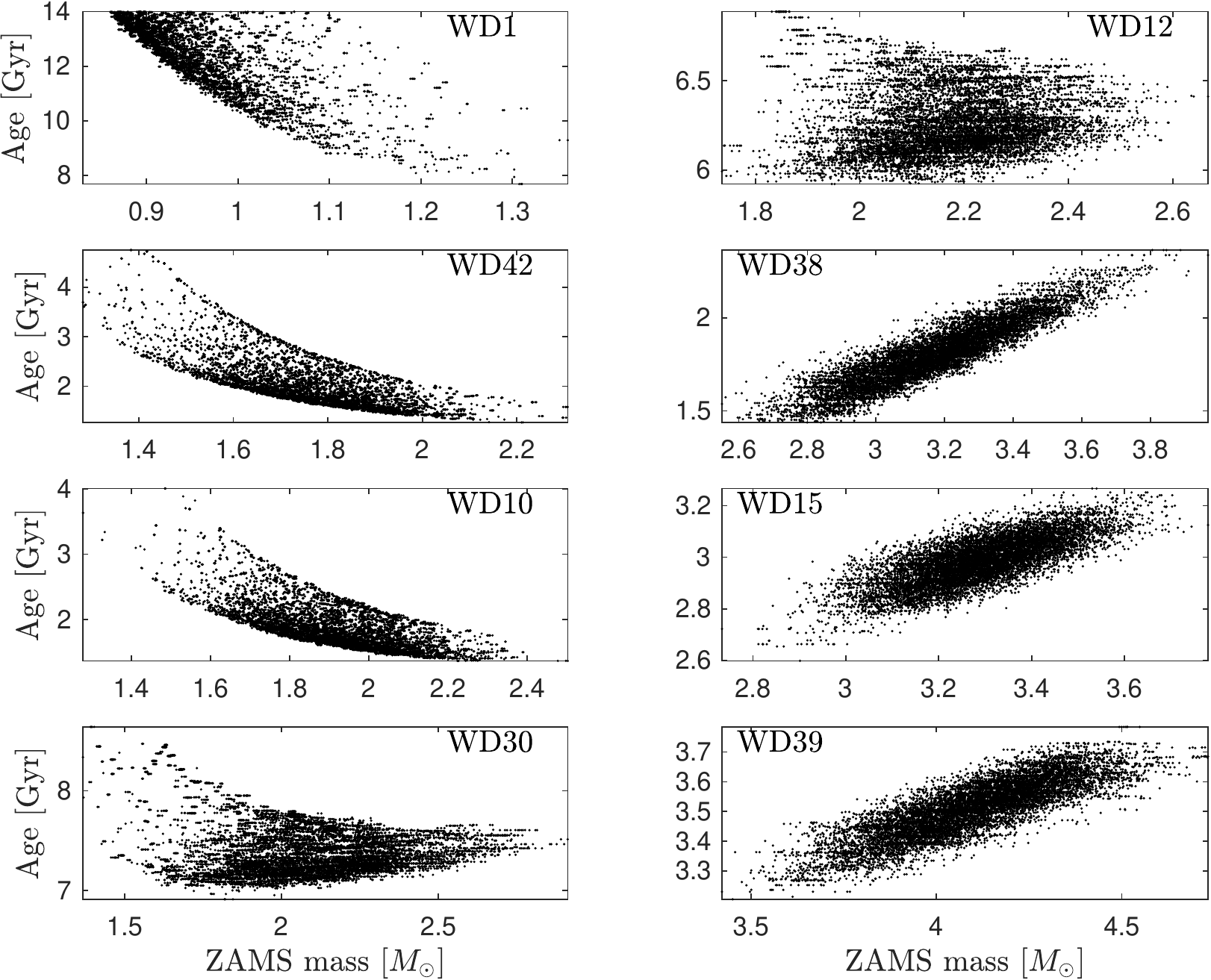} 
  \caption{The joint mass vs. age posterior distribution derived from BASE-9 modeling
  for eight example WD candidates.  The panels list the specific WD and are ordered
by increase ZAMS mass of the WD.  The panels show that there are precise
(though covariant) constraints on both the ages and the precursor mass. 
Note that the first panel presents the case of an overly low inferred WD mass that in practice could not be fit by BASE-9 in a manner
consistent with its parallax.  The only hint of that issue in this
particular diagram is that age is running up against the age of the
Universe.}
\label{fig:fig_example}
\end{figure}

Figure \ref{fig:fig_example} presents the joint posterior distributions ({PDF})
for eight example WDs.  Panels show the zero-age main sequence (ZAMS) mass vs.\
age plane, with each dot presenting a {PDF} sample.  The panels are sorted in
order of increasing mass.
The first panel, for \texttt{WD 1}, shows an example where the parallax prior
mean is {\it inconsistent} with the posterior distance distribution: models
would like to predict a star older than the age of the Universe.  This star is
one of the 15 candidate WDs whose luminosities are above the $0.5\,\msun$ model
in Figure \ref{fig:fig_data_cooling_curve}.  For the other seven WDs presented
Fig.\,\ref{fig:fig_example} and for all but the 15 problematical objects
identified in Figure \ref{fig:fig_data_cooling_curve} (red circles), their
posterior distance distributions are consistent with their companion parallax
prior, indicating that the model star could readily fit the data at 
the appropriate luminosity.  The age precisions among the eight cases in Figure 
\ref{fig:fig_data_cooling_curve} range from $90\,\Myr$ to $1.46\,\Gyr$.  Four of these
eight WDs have fractional age errors of only 3\%, and the WD with the poorest
age constraint (\texttt{WD 42}, with a ZAMS mass of $1.75 \pm 0.15\,\msun$ and age =
$2.1 \pm 0.5$ \Gyr) still provides meaningful age information.  This figure also 
indicates that a more constraining parallax prior, which would in turn further 
constrain the WD mass and thereby its ZAMS mass, would additionally improve the 
age precision for these WDs.  

The formal uncertainties in the fitted WD ages are dominated by the parallax
precision.  While WD models are mature and have benefited from substantial tests
in star clusters, nearby binaries, and asteroseismology, the {\it accuracy} of
the ages may still be poorer than the precision in certain regions of parameter
space.  Particularly WDs with ZAMS masses $\la 2\,\msun$ or WDs with surface
effective temperatures lower than about 5000\,K are challenging.  Gaia
parallaxes tightly constrain the present mass of cool WDs. But when that mass is
mapped back onto the ZAMS, small uncertainties in mass transform to large
uncertainties in the time a WD spent evolving as a MS star.  Additionally, the
IFMR is not known perfectly, and small adjustments in the IFMR may change the
precursor mass values and thus the pre-WD ages, especially for low-mass
precursors.  Thus, for those objects, we can derive a precise cooling age, but
not a precise total age.  For WDs with $\teff \leq 5000$\,K, issues arise both
in our present understanding of their atmospheres and possibly with additional
sources of energy release during crystallization \citep[e.g.][]{Horowitz2010}.
We can avoid most of these problems by focusing on the WDs in a suitable mass
and temperature range.  Nevertheless, formal tests on WD ages have not yet been
performed at the level of the best of these WD age precisions; we will have to
await tests that can be performed in open clusters and WD-WD pairs with Gaia
DR2.  At this point, we would like to emphasize that the WD ages we derive
should be highly precise and deliver excellent {\it relative} ages.  These ages
are likely to be accurate at the 5-10\% level, subject to further testing.

\begin{figure}
\includegraphics[width=\columnwidth]{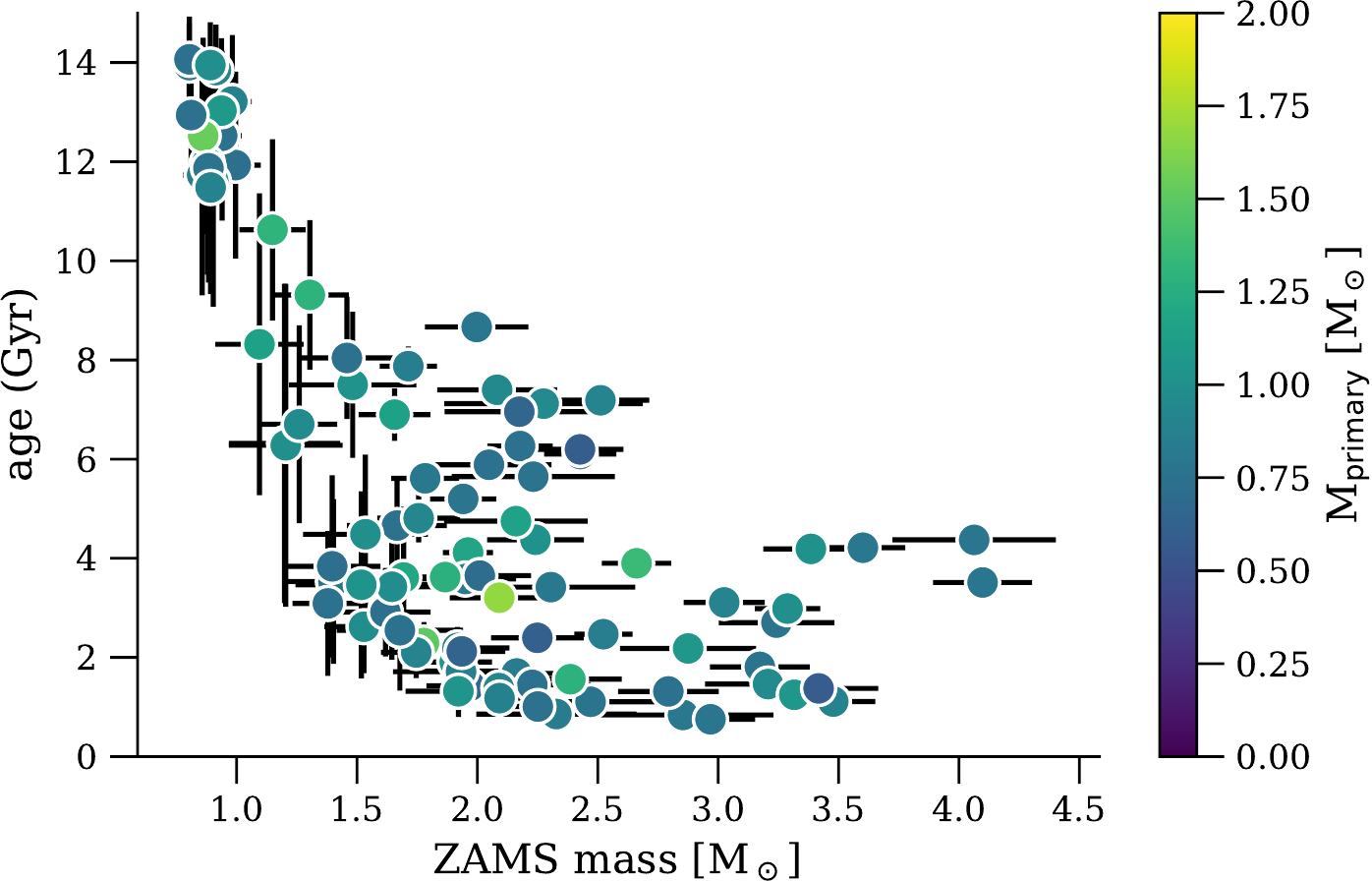} 
\caption{Age in Gyrs vs. ZAMS mass of the precursor in solar units for the 91 WD candidates, all of which are companions to TGAS main sequence stars. Error bars indicate the marginalized $\pm 1\sigma$ age and mass uncertainties;
note that in many cases the age uncertainties are smaller than the symbols.
Age uncertainties for WDs with precursor ZAMS masses $\leq 1.7\,\msun$ are large, because of
the increasing fraction of the system lifetime spent on the MS (rather than as a cooling WD).
The color indicates the mass of the MS primary, showing that most of them are low-mass stars
($M\lesssim 1\,\msun$), whose ages could not be inferred well from isochrones or asteroseismology.
}
\label{fig:fig_mass_age}
\end{figure}

The 91 WDs that BASE-9 fit consistently with the parallaxes are plotted in
Figure \ref{fig:fig_mass_age}.  The error bars represent $\pm$1 standard
deviations in age and ZAMS mass, respectively.  Their colors indicate the
approximative initial mass of their MS companion using their 2MASS photometry
(and the strong prior that they live on the main-sequence).  Age uncertainties
drop rapidly with ZAMS masses $\ge 1.7\,\msun$. The relative age uncertainties, in
the sense $\sigma(\tau_{age})/\langle \tau_{age} \rangle$, range from the highly
precise value of 1.9\% to as poor as 54.5\% at the low ZAMS mass end.  Of these
91 WDs, 42 have relative age precisions better than 10\% and 67 have relative
age precisions of better than 20\%.  The objects plotted in Figure
\ref{fig:fig_mass_age} are both the largest sample of field WDs and the largest
sample of WD - MS pairs with precise ages. 

\section{Discussion and Outlook}\label{sec:Outlook}

In this paper we carried out a pilot study for one of the many applications of
using Gaia data to constrain stellar properties. We identified systems where
Gaia parallaxes gave us distances to nearby ($<200\,\pc$) main sequence stars,
and where common proper motion information from the GPS1 catalog provided strong
evidence for a wide (and equidistant) WD companion. Our analysis nearly doubles
the number of such known wide binaries with parallax distances. 

We applied the BASE-9 modeling to infer ages for the white dwarfs, which must be
the same as those of the MS stars. Achieving better than 10\% age precision for
42 systems, and better than 20\% for another 25 systems (67 in total)
constitutes an order of magnitude increase in the number of low-mass ($\sim
1~N_\odot$) MS field stars for which ages are known with that precision. This
approach seems particularly suited to obtain precise ages for low-mass
($<1\,\msun$) MS stars, where most other methods fail for field stars. The
majority of our systems have ages of 1-8 \Gyr, an age range that is poorly
sampled by known clusters. 

To realize the scientific potential of the sample at hand, spectroscopic
follow-up is necessary in two respects. First, simple low-resolution
spectroscopy needs to verify which of these WDs are actually DA WD's, as assumed
in the modeling. Second, higher-resolution spectroscopy of the bright
($m<11$\,mag) MS stars should be used to determine their detailed abundance
pattern, to increase well-calibrated constraints of $[\vec{X}/Fe]$, i.e.,
$\tau_{age}$ for chemical evolution modeling. We are currently pursuing this
follow-up.

While this particular sample will of course be superseded by the data from
Gaia's DR2 (in April 2018), this overall approach will be particularly powerful
in light of the full Gaia data.  For studying the WD's themselves, precise
parallaxes will be paramount, especially for the oldest and faintest WDs. In
these case, the boost in parallax precision transferred from the MS star, will
aid the analysis. In turn, identifying WD companions to MS stars mostly by their
common proper motion, will greatly enlarge the volumes over which this analysis
can be done (compared to insisting on precise parallaxes for both the MS and the
WD).

\acknowledgments
This project was developed in part at the 2017 Heidelberg Gaia Sprint, hosted by
the Max-Planck-Institut f\"ur Astronomie, Heidelberg.

This work has made use of data from the European Space Agency (ESA) mission Gaia
(http://www.cosmos.esa.int/gaia), processed by the Gaia Data Processing and
Analysis Consortium (DPAC, http://www.cosmos.esa.int/web/gaia/dpac/consortium).
Funding for the DPAC has been provided by national institutions, in particular
the institutions participating in the Gaia Multilateral Agreement.

Parts of the work has used the German Astrophysical Virtual Observatory (GAVO)
team at the Zentrum für Astronomie Heidelberg. GAVO is funded through the
Verbundforschung of the German Ministry for Research (BMBF).

H.-W.R.'s research contribution is supported by the European Research Council
under the European Union's Seventh Framework Programme (FP 7) ERC Grant
Agreement n.$\,$[321035].  T.v.H.'s research contribution is supported by the
National Science Foundation Award AST-1715718.

H.-W.R. gratefully acknowledges early discussion with Dan Maoz that proved
seminal for this paper. 

\bibliographystyle{apj}
\bibliography{tc.bib}

\begin{thebibliography}{24}
\expandafter\ifx\csname natexlab\endcsname\relax\def\natexlab#1{#1}\fi

\bibitem[{{Althaus} \& {Benvenuto}(1998)}]{Althaus1998}
{Althaus}, L.~G., \& {Benvenuto}, O.~G. 1998, \mnras, 296, 206

\bibitem[{{Angus} {et~al.}(2015){Angus}, {Aigrain}, {Foreman-Mackey}, \&
  {McQuillan}}]{Angus2015}
{Angus}, R., {Aigrain}, S., {Foreman-Mackey}, D., \& {McQuillan}, A. 2015,
  \mnras, 450, 1787

\bibitem[{{Bergeron} {et~al.}(2001){Bergeron}, {Leggett}, \&
  {Ruiz}}]{Bergeron2001}
{Bergeron}, P., {Leggett}, S.~K., \& {Ruiz}, M.~T. 2001, \apjs, 133, 413

\bibitem[{{Bergeron} {et~al.}(1995){Bergeron}, {Wesemael}, \&
  {Beauchamp}}]{Bergeron1995}
{Bergeron}, P., {Wesemael}, F., \& {Beauchamp}, A. 1995, \pasp, 107, 1047

\bibitem[{{Chaplin} {et~al.}(2014){Chaplin}, {Basu}, {Huber}, {Serenelli},
  {Casagrande}, {Silva Aguirre}, {Ball}, {Creevey}, {Gizon}, {Handberg},
  {Karoff}, {Lutz}, {Marques}, {Miglio}, {Stello}, {Suran}, {Pricopi},
  {Metcalfe}, {Monteiro}, {Molenda-{\.Z}akowicz}, {Appourchaux},
  {Christensen-Dalsgaard}, {Elsworth}, {Garc{\'{\i}}a}, {Houdek}, {Kjeldsen},
  {Bonanno}, {Campante}, {Corsaro}, {Gaulme}, {Hekker}, {Mathur}, {Mosser},
  {R{\'e}gulo}, \& {Salabert}}]{Chaplin2014}
{Chaplin}, W.~J., {Basu}, S., {Huber}, D., {Serenelli}, A., {Casagrande}, L.,
  {Silva Aguirre}, V., {Ball}, W.~H., {Creevey}, O.~L., {Gizon}, L.,
  {Handberg}, R., {Karoff}, C., {Lutz}, R., {Marques}, J.~P., {Miglio}, A.,
  {Stello}, D., {Suran}, M.~D., {Pricopi}, D., {Metcalfe}, T.~S., {Monteiro},
  M.~J.~P.~F.~G., {Molenda-{\.Z}akowicz}, J., {Appourchaux}, T.,
  {Christensen-Dalsgaard}, J., {Elsworth}, Y., {Garc{\'{\i}}a}, R.~A.,
  {Houdek}, G., {Kjeldsen}, H., {Bonanno}, A., {Campante}, T.~L., {Corsaro},
  E., {Gaulme}, P., {Hekker}, S., {Mathur}, S., {Mosser}, B., {R{\'e}gulo}, C.,
  \& {Salabert}, D. 2014, \apjs, 210, 1

\bibitem[{{De Gennaro} {et~al.}(2008){De Gennaro}, {von Hippel}, {Winget},
  {Kepler}, {Nitta}, {Koester}, \& {Althaus}}]{DeGennaro2008}
{De Gennaro}, S., {von Hippel}, T., {Winget}, D.~E., {Kepler}, S.~O., {Nitta},
  A., {Koester}, D., \& {Althaus}, L. 2008, \aj, 135, 1

\bibitem[{{Dotter}(2016)}]{Dotter2016}
{Dotter}, A. 2016, \apjs, 222, 8

\bibitem[{{Dotter} {et~al.}(2008){Dotter}, {Chaboyer}, {Jevremovi{\'c}},
  {Kostov}, {Baron}, \& {Ferguson}}]{dotter2008}
{Dotter}, A., {Chaboyer}, B., {Jevremovi{\'c}}, D., {Kostov}, V., {Baron}, E.,
  \& {Ferguson}, J.~W. 2008, \apjs, 178, 89

\bibitem[{{Gaia Collaboration} {et~al.}(2016){Gaia Collaboration}, {Brown},
  {Vallenari}, {Prusti}, {de Bruijne}, {Mignard}, {Drimmel}, {Babusiaux},
  {Bailer-Jones}, {Bastian}, \& et~al.}]{Gaia2016}
{Gaia Collaboration}, {Brown}, A.~G.~A., {Vallenari}, A., {Prusti}, T., {de
  Bruijne}, J.~H.~J., {Mignard}, F., {Drimmel}, R., {Babusiaux}, C.,
  {Bailer-Jones}, C.~A.~L., {Bastian}, U., \& et~al. 2016, \aap, 595, A2

\bibitem[{{Horowitz} {et~al.}(2010){Horowitz}, {Schneider}, \&
  {Berry}}]{Horowitz2010}
{Horowitz}, C.~J., {Schneider}, A.~S., \& {Berry}, D.~K. 2010, Physical Review
  Letters, 104, 231101

\bibitem[{{Montgomery} {et~al.}(1999){Montgomery}, {Klumpe}, {Winget}, \&
  {Wood}}]{Montgomery1999}
{Montgomery}, M.~H., {Klumpe}, E.~W., {Winget}, D.~E., \& {Wood}, M.~A. 1999,
  \apj, 525, 482

\bibitem[{{O'Malley} {et~al.}(2013){O'Malley}, {von Hippel}, \& {van
  Dyk}}]{OMalley2013}
{O'Malley}, E.~M., {von Hippel}, T., \& {van Dyk}, D.~A. 2013, \apj, 775, 1

\bibitem[{{Renedo} {et~al.}(2010){Renedo}, {Althaus}, {Miller Bertolami},
  {Romero}, {C{\'o}rsico}, {Rohrmann}, \& {Garc{\'{\i}}a-Berro}}]{Renedo2010}
{Renedo}, I., {Althaus}, L.~G., {Miller Bertolami}, M.~M., {Romero}, A.~D.,
  {C{\'o}rsico}, A.~H., {Rohrmann}, R.~D., \& {Garc{\'{\i}}a-Berro}, E. 2010,
  \apj, 717, 183

\bibitem[{{Salaris} {et~al.}(2009){Salaris}, {Serenelli}, {Weiss}, \& {Miller
  Bertolami}}]{Salaris2009}
{Salaris}, M., {Serenelli}, A., {Weiss}, A., \& {Miller Bertolami}, M. 2009,
  \apj, 692, 1013

\bibitem[{{Soderblom}(2010)}]{soderblom2010}
{Soderblom}, D.~R. 2010, \araa, 48, 581

\bibitem[{{Stein} {et~al.}(2013){Stein}, {van Dyk}, {von Hippel}, {DeGennaro},
  {Jeffery}, \& {Jefferys}}]{Stein2013}
{Stein}, N.~M., {van Dyk}, D.~A., {von Hippel}, T., {DeGennaro}, S., {Jeffery},
  E.~J., \& {Jefferys}, W.~H. 2013, Statistical Analysis and Data Mining: The
  ASA Data Science Journal, Vol.~9, Issue 1, p.~34-52, 6, 34

\bibitem[{{Stenning} {et~al.}(2016){Stenning}, {Wagner-Kaiser}, {Robinson},
  {van Dyk}, {von Hippel}, {Sarajedini}, \& {Stein}}]{Stenning2016}
{Stenning}, D.~C., {Wagner-Kaiser}, R., {Robinson}, E., {van Dyk}, D.~A., {von
  Hippel}, T., {Sarajedini}, A., \& {Stein}, N. 2016, \apj, 826, 41

\bibitem[{{Tian} {et~al.}(2017){Tian}, {Gupta}, {Sesar}, {Rix}, {Martin},
  {Liu}, {Goldman}, {Platais}, {Kudritzki}, \& {Waters}}]{Tian2017}
{Tian}, H.-J., {Gupta}, P., {Sesar}, B., {Rix}, H.-W., {Martin}, N.~F., {Liu},
  C., {Goldman}, B., {Platais}, I., {Kudritzki}, R.-P., \& {Waters}, C.~Z.
  2017, \apjs, 232, 4

\bibitem[{{Torres} {et~al.}(2010){Torres}, {Andersen}, \&
  {Gim{\'e}nez}}]{Torres2010}
{Torres}, G., {Andersen}, J., \& {Gim{\'e}nez}, A. 2010, \aapr, 18, 67

\bibitem[{{Tremblay} \& {Bergeron}(2008)}]{Tremblay2008}
{Tremblay}, P.-E., \& {Bergeron}, P. 2008, \apj, 672, 1144

\bibitem[{{Tremblay} {et~al.}(2017){Tremblay}, {Gentile-Fusillo}, {Raddi},
  {Jordan}, {Besson}, {G{\"a}nsicke}, {Parsons}, {Koester}, {Marsh}, {Bohlin},
  {Kalirai}, \& {Deustua}}]{Tremblay2017}
{Tremblay}, P.-E., {Gentile-Fusillo}, N., {Raddi}, R., {Jordan}, S., {Besson},
  C., {G{\"a}nsicke}, B.~T., {Parsons}, S.~G., {Koester}, D., {Marsh}, T.,
  {Bohlin}, R., {Kalirai}, J., \& {Deustua}, S. 2017, \mnras, 465, 2849

\bibitem[{{van Dyk} {et~al.}(2009){van Dyk}, {Degennaro}, {Stein}, {Jefferys},
  \& {von Hippel}}]{vanDyk2009}
{van Dyk}, D.~A., {Degennaro}, S., {Stein}, N., {Jefferys}, W.~H., \& {von
  Hippel}, T. 2009, Annals of Applied Statistics, 3, 117

\bibitem[{{von Hippel} {et~al.}(2006){von Hippel}, {Jefferys}, {Scott},
  {Stein}, {Winget}, {De Gennaro}, {Dam}, \& {Jeffery}}]{vonHippel2006}
{von Hippel}, T., {Jefferys}, W.~H., {Scott}, J., {Stein}, N., {Winget}, D.~E.,
  {De Gennaro}, S., {Dam}, A., \& {Jeffery}, E. 2006, \apj, 645, 1436

\bibitem[{{Williams} {et~al.}(2009){Williams}, {Bolte}, \&
  {Koester}}]{williams2009}
{Williams}, K.~A., {Bolte}, M., \& {Koester}, D. 2009, \apj, 693, 355

\end{thebibliography}

\appendix

\section{GPS1$\times$TGAS Query}
\label{appendix:gps1query}

In this section, we detail the selection query we performed on TGAS and GPS1 catalogs.

Matching GPS1 against TGAS will report all the stars from GPS1 within some radius that could potentially be associated with a TGAS bright star.
If we also filter on parallax and motion similarity this will only give co-moving pairs.
We consider nearby objects according to TGAS parallaxes as
\begin{equation}\label{eq:distance}
\begin{split}
\textrm{distance}((\alpha, \delta)_{GPS1}, &(\alpha, \delta)_{TGAS})\,[\rm deg] \\
& < 10.3 \times \frac{\varpi [\mas]}{3600}.
\end{split}
\end{equation}
Further tuning can be done by adding a contamination model, though this is out of the proof-of-concept scope of this paper.
In addition, we need to only conserve good parallaxes within a $200\,\pc$ ($5\,\mas$) volume around the Sun as
\begin{equation}\label{eq:good_parallax}
\begin{split}
& \varpi \geq 5 \, \mas \\
& \frac{\varpi}{\sigma_\varpi} > 20,
\end{split}
\end{equation}
and relatively good motion precision in GPS1
\begin{equation}\label{eq:good_motion}
\sqrt{\sigma^2_{\mu,\alpha} + \sigma^2_{\mu,\delta}} < 6\,\masyr
\end{equation}

Additionally, we want pairs of objects that are co-moving according to both surveys (within their uncertainties). Therefore we select pairs that appear co-moving within $5-\sigma$ uncertainties:
\begin{equation}
\begin{split}
      & \frac{\left((\mu_\alpha^\star)_{GPS1} - (\mu_\alpha^\star)_{TGAS}\right)^2}{\left((\sigma_{\mu,\alpha})_{GPS1}^2 + (\sigma_{\mu,\alpha})_{TGAS}^2  \right)}\\
\quad & +
\frac{\left((\mu_\delta)_{GPS1} - (\mu_\delta)_{TGAS}\right)^2}{\left((\sigma_{\mu,\delta})_{GPS1}^2 + (\sigma_{\mu,\delta})_{TGAS}^2  \right)}
\leq (5 \,\masyr) ^2.
\end{split}
\end{equation}

However, many objects with small motion where actually contaminant or main-sequence objects. Therefore we also include a revised cut that rejects objects with small motions (despite leading to incompleteness):
\begin{equation}
\begin{split}
      & \sqrt{(\mu_\alpha^\star)_{TGAS}^2 + (\mu_\delta)_{TGAS}^2}~[\masyr] \\
\quad & > {25}\left(\frac{1000}{0.3\,\varpi [\mas]} \times \textrm{distance}((\alpha, \delta)_{GPS1}, (\alpha, \delta)_{TGAS})\right)^{0.7}.
\end{split}
\end{equation}
Note that the constant and power of the above equation are results of an empirical inspection. 
Finally, we also added color terms that avoid having main-sequence objects and we also select good photometry for their SED analysis. Based on empirical definitions we added the following selections:

\begin{equation}
\begin{split}
& \left|(g - i) -1.6 \times (g - r) + 0.1 \right| < 0.15\, \mathrm{mag},\\
& \left(\sigma_g,\, \sigma_r ,\, \sigma_i,\, \sigma_z\right) < 0.05\, \mathrm{mag},
\end{split}
\end{equation}

This selection translates into the following ADQL query. As GAVO is currently the only service providing the GPS1 catalog, the field names correspond to their definition, and may vary when using other sources (e.g., VizieR, Gaia Archive).

\scriptsize{

\begin{verbatim}
SELECT 
    db.obj_id, db.ra, db.dec, db.e_ra, db.e_dec, db.pmra, db.e_pmra, 
    db.pmde, db.e_pmde, db.magg,  db.e_magg, db.magr,  db.e_magr, 
    db.magi, db.e_magi, db.magz, db.e_magz, db.magy, db.e_magy, 
    db.magj, db.e_magj, db.magh, db.e_magh, db.magk, db.e_magk, 
    db.maggaia, db.e_maggaia,  tc.source_id, tc.ra, tc.dec, 
    tc.ra_error, tc.dec_error, tc.l, tc.b, tc.pmra, tc.pmdec, 
    tc.pmra_error, tc.pmdec_error, tc.parallax, tc.parallax_error,
    tc.phot_g_mean_mag,  tc.phot_variable_flag, 
    tc.astrometric_excess_noise_sig, tc.ra_dec_corr, tc.ra_pmra_corr,
    tc.ra_pmdec_corr, tc.dec_pmra_corr, tc.dec_pmdec_corr, 
    tc.pmra_pmdec_corr, tc.ra_parallax_corr, tc.dec_parallax_corr, 
    tc.parallax_pmra_corr, tc.parallax_pmdec_corr, tc.phot_g_n_obs,
    distance(POINT(’icrs’, db.ra, db.dec),  
             POINT(’icrs’, tc.ra, tc.dec)) AS pairdistance
FROM tgas.main AS tc
JOIN gps1.main AS db
ON
    1 = contains(POINT(’ICRS’, db.ra, db.dec), 
                 CIRCLE(’ICRS’, tc.ra, tc.dec, 10.3 * tc.parallax/3600.))
WHERE
    parallax >= 5 AND parallax / parallax_error > 20
AND
    (power((db.pmra * 3.6 * 1e6 - tc.pmra), 2) / 
        (power(db.e_pmra * 3.6 * 1e6 ,2) + power(tc.pmra_error, 2))  +
        power((db.pmde * 3.6 * 1e6 - tc.pmdec), 2) / 
            (power(db.e_pmde * 3.6 * 1e6,2) + power(tc.pmdec_error, 2))
     ) < 25
AND
    sqrt((power(tc.pmra, 2)+ power(tc.pmdec, 2) )) > 
      25 *  power(distance(POINT('icrs', db.ra, db.dec), 
                           POINT('icrs', tc.ra, tc.dec)) 
                  * (100./tc.parallax) / 0.03, 0.7)
AND
    db.e_magg < 0.05 AND  db.e_magr < 0.05
AND
    db.e_magi < 0.05 AND db.e_magz < 0.05
AND
    abs((magg - magi) - 1.6*(magg - magr)+0.1) < 0.15
\end{verbatim}
}

Note that on Fig.\ref{fig:fig_selection}, the red selection corresponds to this query, while the blue selection results from the same query were we only apply the \verb|JOIN| and the two first \verb|WHERE| conditions.

\section{Catalogs}

In this section we describe the content of the catalog generated during this study.

The catalog contains the photometric and astrometric data for all of the WD candidates of this study. For each star, we also provide the mean, median and standard deviation of the posterior PDF of the WD properties, esp. age and ZAMS mass.
In addition, the catalog contains the matched MS component 2MASS (J, H, K), and WISE (W1, W2, W3, W4) photometry as well as our mass estimates and uncertainties.

\begin{table*}
\caption{Catalog column description}
\scriptsize
\begin{tabular}{rcl|rcl}
\hline
Column & Units & Description & Column & Units & Description\\\hline\hline
  \texttt{source\_id} & & Gaia DR1 identifier &                                       \texttt{AllWISE} &  &  AllWise identifier \\
  \texttt{magg} & mag & Gaia DR1 $G$ magnitude (of the WD) &                          \texttt{gps1\_ra} & deg &  right ascension from GPS1 \\
  \texttt{e\_magg} & mag & Gaia $G$ magnitude uncertainty &                           \texttt{gps1\_dec} & deg & declination from GPS1\\
  \texttt{magr} & mag & GPS1 $r$ magnitude &                                          \texttt{gps1\_e\_ra} & deg & GPS1 RA uncertainty \\
  \texttt{e\_magr} & mag & GPS1 $r$ uncertainty &                                     \texttt{gps1\_e\_dec} & deg & GPS1 DEC uncertainty \\
  \texttt{magi} & mag & GPS1 $i$ magnitude &                                          \texttt{gps1\_pmra} & deg/yr$^{-1}$ &  GPS1 $\mu_\alpha^\star$\\
  \texttt{e\_magi} & mag & GPS1 $i$ uncertainty &                                     \texttt{gps1\_pmde} & deg/yr$^{-1}$ &  GPS1 $\mu_\delta$\\
  \texttt{magz} & mag & GPS1 $z$ magnitude &                                          \texttt{gps1\_e\_pmde} &  deg/yr$^{-1}$ &  GPS1 $\mu_\alpha^\star$ uncertainty\\
  \texttt{e\_magz} & mag & GPS1 $z$ uncertainty &                                     \texttt{gps1\_e\_pmra} & deg/yr$^{-1}$ &  GPS1 $\mu_\delta$ uncertainty\\
  \texttt{magy} & mag & GPS1 $y$ magnitude &                                          \texttt{primary\_Hmag} & mag &  primary $H$ photometry\\
  \texttt{e\_magy} & mag & GPS1 $y$ uncertainty &                                     \texttt{primary\_Jmag} & mag &  primary $J$ photometry\\
  \texttt{magj} & mag & GPS1 $J$ magnitude &                                          \texttt{primary\_Kmag} & mag &  primary $K$ photometry\\
  \texttt{e\_magj} & mag & GPS1 $J$ uncertainty &                                     \texttt{primary\_W1mag} & mag & primary $W1$ photometry \\
  \texttt{magh} & mag & GPS1 $H$ magnitude &                                          \texttt{primary\_W2mag} & mag & primary $W2$ photometry \\
  \texttt{e\_magh} & mag & GPS1 $H$ uncertainty &                                     \texttt{primary\_W3mag} & mag & primary $W3$ photometry \\
  \texttt{magk} & mag & GPS1 $K$ magnitude &                                          \texttt{primary\_W4mag} & mag & primary $W4$ photometry \\
  \texttt{e\_magk} & mag & GPS1 $K$ uncertainty &                                     \texttt{primary\_e\_Hmag} & mag &  primary $H$ uncertainty\\
  \texttt{maggaia} & mag & GPS1 Gaia $G$ magnitude &                                  \texttt{primary\_e\_Jmag} & mag &  primary $J$ uncertainty\\
  \texttt{e\_maggaia} & mag & GPS1 converted Gaia $G$ uncertainty &                   \texttt{primary\_e\_Kmag} & mag &  primary $K$ uncertainty\\
  \texttt{parallax} & \mas & Gaia DR1 Parallax (Primary) &                            \texttt{primary\_e\_W1mag} & mag & primary $W1$ uncertainty\\
  \texttt{parallax\_error} & \mas & Gaia DR1 parallax uncertainty &                   \texttt{primary\_e\_W2mag} & mag & primary $W2$ uncertainty\\
  \texttt{mn\_Age} & $\Gyr$ & posterior mean WD age &                                 \texttt{primary\_e\_W3mag} & mag & primary $W3$ uncertainty\\
  \texttt{mn\_fe} & dex & posterior mean $\feh$ &                                     \texttt{primary\_e\_W4mag} & mag & primary $W4$ uncertainty\\
  \texttt{mn\_mod} & mag & posterior mean distance modulus &                          \texttt{primary\_mass\_p16} & \msun &  16th mass percentile\\
  \texttt{mn\_mass} & \msun & posterior mean WD mass &                                \texttt{primary\_mass\_p50} & \msun &  50th mass percentile\\
  \texttt{mn\_cAge} & \Gyr & posterior mean WD cooling age&                                \texttt{primary\_mass\_p84} & \msun &  84th mass percentile\\
  \texttt{mn\_pAge} & \Gyr & posterior mean WD precusor's age &                            \texttt{tgas\_ra} &  deg &  right ascension from TGAS\\
  \texttt{md\_Age} & $\Gyr$ & posterior median WD age &                               \texttt{tgas\_ra\_error} &  \mas &  TGAS right ascension uncertainty \\
  \texttt{md\_fe} & dex & posterior median $\feh$ &                                   \texttt{tgas\_dec} & deg &  declination from TGAS\\
  \texttt{md\_mod} & mag & posterior median distance modulus &                        \texttt{tgas\_dec\_error} & \mas &  TGAS declination uncertainty\\
  \texttt{md\_mass} & \msun & posterior median WD mass &                              \texttt{tgas\_b} &  deg &  Galactic latitude from TGAS\\
  \texttt{md\_cAge} & \Gyr & posterior median WD cooling age&                                \texttt{tgas\_l} & deg &  Galactic longitude from TGAS\\
  \texttt{md\_pAge} & \Gyr & posterior median WD precusor's age &                            \texttt{tgas\_Gmag} & mag &  primary TGAS G magnitude\\
  \texttt{st\_Age} & $\Gyr$ & posterior standard deviation WD age &                   \texttt{tgas\_pmdec} &  $\masyr$ &  TGAS $\mu_\alpha^\star$\\
  \texttt{st\_fe} & dex & posterior standard deviation $\feh$ &                       \texttt{tgas\_pmdec\_error} &  $\masyr$ &  TGAS $\mu_\alpha^\star$ uncertainty\\
  \texttt{st\_mod} & mag & posterior standard deviation distance modulus &            \texttt{tgas\_pmra} & $\masyr$ &  TGAS $\mu_\delta$\\
  \texttt{st\_mass} & \msun & posterior standard deviation WD mass &                  \texttt{tgas\_pmra\_error} &  $\masyr$ & TGAS $\mu_\delta$ uncertainty\\
  \texttt{st\_cAge} & \Gyr & posterior standard deviation WD cooling age &                               \texttt{primary\_source\_id} & & primary TGAS DR1 identifier \\
  \texttt{st\_pAge} & \Gyr & posterior standard deviation WD precusor's age &    & & \\
\hline
\end{tabular}
\end{table*}


\end{document}